\documentclass[twocolumn,12pt,cites,graphicx]{article}
\usepackage{epsfig}

\title{Confinement Effects on Phase Behavior of Soft Matter Systems}
\author{Kurt Binder\thanks{Institut f\"ur Physik, Johannes Gutenberg Universit\"at Mainz,
Staudinger Weg 7, 55099 Mainz, Germany. E-mail: kurt.binder@uni-mainz.de} \and
J\"urgen Horbach\thanks{Institut f\"ur Materialphysik im Weltraum, Deutsches Zentrum f\"ur Luft- und
Raumfahrt (DLR), 51170 K\"oln, Germany. E-mail: juergen.horbach@dlr.de} \and
Richard Vink\thanks{Institut f\"ur Theoretische Physik, Georg-August Universit\"at,
Friedrich-Hund-Platz 1, 37077 G\"ottingen, Germany. E-mail:vink@theorie.physik.uni-goettingen.de} \and 
Andres De Virgiliis\thanks{Instituto de Investigaciones Fisicoquimicas, UNLP, CONICET,
Sucursal 4, Casilla de Correo 16, 1900 La Plata, Argentina. E-mail: adevir@inifta.unlp.edu.ar}}

\begin{document}

\maketitle
\renewcommand{\thefootnote}{\fnsymbol{footnote}}

\noindent When systems that can undergo phase separation between
two coexisting phases in the bulk are confined in thin film
geometry between parallel walls, the phase behavior can be
profoundly modified. These phenomena shall be described and
exemplified by computer simulations of the Asakura-Oosawa model
for colloid-polymer mixtures, but applications to other soft
matter systems (e.g.~confined polymer blends) will also be
mentioned. Typically a wall will prefer one of the phases, and
hence the composition of the system in the direction perpendicular
to the walls will not be homogeneous. If both walls are of the
same kind, this effect leads to a distortion of the phase diagram
of the system in thin film geometry, in comparison with the bulk,
analogous to the phenomenon of ``capillary condensation'' of
simple fluids in thin capillaries. In the case of ``competing
walls'', where both walls prefer different phases of the two
phases coexisting in the bulk, a state with an interface parallel
to the walls gets stabilized. The transition from the disordered
phase to this ``soft mode phase'' is rounded by the finite
thickness of the film and not a sharp phase transition. However, a
sharp transition can occur where this interface gets localized at
(one of) the walls. The relation of this interface localization
transition to wetting phenomena is discussed. Finally, an outlook
to related phenomena is given, such as the effects of confinement
in cylindrical pores on the phase behavior, and more complicated
ordering phenomena (lamellar mesophases of block copolymers or
nematic phases of liquid crystals under confinement).

\section{Introduction}
The current interest in the construction of nanoscopic devices
\cite{1,2,3,4,5} demands a better understanding of the phase
behavior of fluids confined in pores or slits of nanoscopic linear
dimensions \cite{6,7,8,9,10,11,12}. Knowledge on the phase
behavior of confined fluids is a prerequisite to understand their
dynamics \cite{13,14,15}, as well as for the analysis of flow
through very thin capillaries \cite{16,17}, nanoscale capillary
imbibition \cite{18,19}, and related microfluidic or nanofluidic
devices.

Obviously, an interplay must be expected between surface effects
on the fluid due to the confining walls, such as adsorption
\cite{20,21,22,23}, formation of wetting (or drying) layers
\cite{24,25,26,27}, and finite size effects \cite{28,29,30} due to
the finite width of the capillary. However, understanding the
nanoscopic confinement of real fluids consisting of small
molecules is very difficult due to additional effects, resulting
from the lateral variation of the wall potential caused by wall
roughness or even the atomistic corrugation \cite{31,32} of the
wall.

While there has been an enormous activity to study theoretically
and by computer simulation confinement effects on simple fluid
models such as the Ising lattice gas model \cite{11} or simple
Lennard-Jones systems \cite{10,12} and rich predictions from
phenomenological theories are available as well
\cite{7,8,9,10,11,12,33,34,35,36,37,38,39}, it is difficult to
find pertinent experiments to which such work could be compared.
However, it is much more promising to study confinement effects on
soft matter systems: due to the mesoscopic length scales of the
particles that one encounters when one studies mixtures of
polymers and colloids \cite{40} or polymer blends \cite{41,42},
effects due to the atomistic corrugation of the
walls are much less important; also the large size of colloidal
particles enables more detailed experimental observations; e.g.
individual particles can be tracked though real space in real time
using confocal microscopy \cite{43} and interface fluctuations in
mixtures of colloids and polymers can be directly observed
\cite{44,45}. Moreover, colloids are model systems for the study
of phase behavior, since by changing suitable parameters the
strength and range of effective interactions can be varied over a
wide range \cite{46,47,48}. Also the interaction of colloidal
particles with the confining walls can be tuned, e.g. by coating
the wall with a polymer brush \cite{49,50,51,52} and controlling
the polymer-wall interaction via variation of the grafting density
and/or chain length of the anchoring flexible polymer
\cite{52,53}. In particular, for colloid-polymer mixtures both the
radius of the (spherical) colloidal particles and the size ratio
between colloids and polymer coils controls the location of the
critical point where the phase separation in a colloid-rich and a
polymer-rich phase sets in \cite{40}.

Similarly, also blends of long flexible polymers are a very
suitable model system to study the effect of confinement in a thin
film geometry on phase separation experimentally as well
\cite{41,42,54}. Again, already the location of the critical
temperature of phase separation in the bulk can be varied over a
wide range, by suitable choices of the polymeric species, and
their chain lengths \cite{55,56,57}. In addition, characteristic
lengths of the problem such as the correlation length of
composition fluctuations \cite{57}, the (intrinsic) interfacial
width between coexisting phases \cite{57}, etc., are much larger
than interatomic distances, and hence also for these systems
experimental probes are available which would lack sensitivity for
small molecule systems. E.g., for the study of the anomalous
broadening of interfaces depending on the film thickness
\cite{41,42}, which is one of the characteristic signatures of the
``soft mode phase'' \cite{36,37} in a system with ``competing
walls'' \cite{38}, a nuclear-reaction based depth
profiling method \cite{54} was used. This method can resolve the
very wide interfaces in such soft matter systems, while it would
be unsuitable for the much narrower interfaces in mixtures of
small molecules.

As is evident from this introductory discussion, and the extensive
literature that has already been quoted, the subject is extremely
rich, and comprehensive coverage could fill a whole book.
Therefore the scope of the present review necessarily must be more
narrow. We shall focus in this review almost exclusively on
confinement effects of colloid-polymer mixtures
\cite{58,59,60,61,62,63}. Only spherical
colloidal particles shall be discussed, although related phenomena can be studied
also for mixtures of polymers with rod-like colloids \cite{64}.
Although extensive work has been done for models of polymer
blends, both using the self-consistent field theory
\cite{65,66,67} and simulations \cite{65,68,69,70}, we shall not
consider this work here, but draw attention to recent reviews
\cite{71,72}. Also, we shall not attempt to review the theory of
wetting phenomena \cite{24,25,26,27} and scaling theories of
capillary condensation \cite{11,33} and interface localization
transitions \cite{36,37,38,70} but rather refer the reader to
another thorough review \cite{11}.

In Sec.~2, we shall briefly recall work on phase behavior of the
model of Asakura and Oosawa (AO) \cite{73} and Vrij \cite{74},
where colloids simply are described as hard spheres which may
neither overlap with each other nor overlap with polymers, while the
latter may overlap with each other with no energy cost, in the bulk
\cite{75,76,77,78,79,80,81,82,83}. In Sec. 3 we shall discuss the
phase behavior of this AO model when it is confined
\cite{58,59,60,61,62} between symmetrical walls a distance $D$
apart, paying attention to the shift of the critical point as a
function of film thickness $D$, and to the change of the critical
behavior. Sec. 4 then describes the behavior encountered for
asymmetric walls \cite{63}, where it is also shown that by
variation of the conditions at the walls one can gradually
crossover from this interface localization transition to a
transition which is of capillary condensation type. Sec. 5 then
presents a summary of the results reviewed here, and gives an
outlook on related findings in other systems, as well as to more
complicated phenomena where the order parameter characterizing the
transition is not a simple scalar quantity (as it is for
gas-liquid or liquid-liquid type phase separation).

\section{Liquid-liquid demixing for the Asakura-Oosawa (AO) 
model in the bulk}
\begin{figure}
\begin{center}
\includegraphics[width=0.45\textwidth]{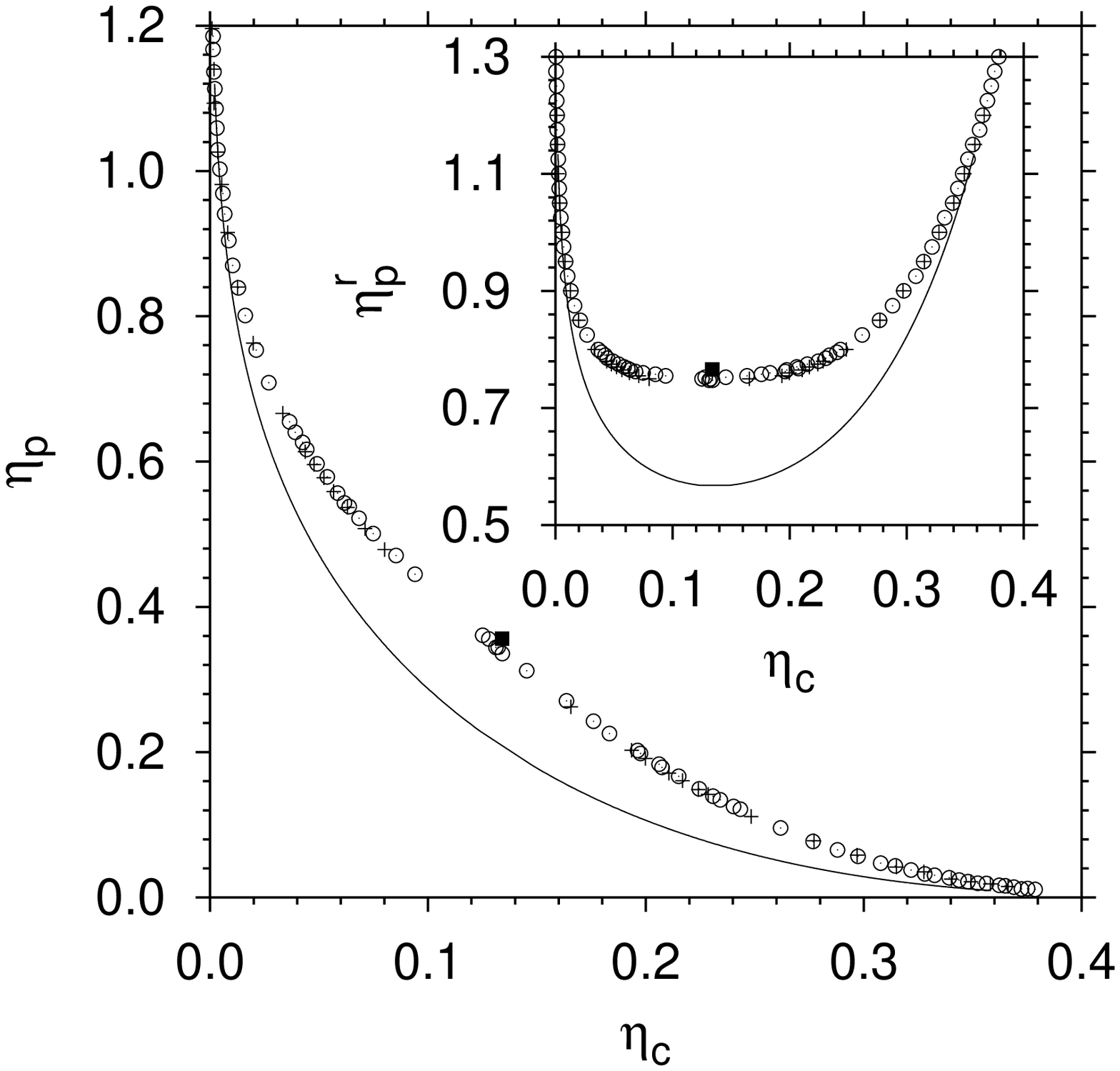}
\caption{Phase diagram of the AO model with a size ratio
$q=R_{\rm p}/R_{\rm c}=0.8$ plotted in the $(\eta_{\rm p}, \eta_{\rm c})$ plane, showing
the coexistence curve for phase separation into a polymer-rich
phase (left) and a colloid-rich phase (right), according to Monte
Carlo (circles) and the free volume mean-field theory of
Lekkerkerker et al.~(full line) \cite{75}. The Monte Carlo data were obtained
using a simulation box with linear dimensions $L_x=L_y=16.7$,
$L_z=33.4$ (some data for a smaller box with $L_x=L_y=13.3$,
$L_z=26.5$ are also included as crosses). The solid square shows the
estimate for the critical point as obtained from a finite size
scaling analysis of the Monte Carlo data, cf.~text. The insert
shows the same data, in the so-called ``reservoir representation''
where $\eta^{\rm r}_{\rm p}$ rather than $\eta_{\rm p}$ is used as a variable. The
one-phase region where colloids and polymers are fully miscible is
always shown in the lower parts of these diagrams, below the
coexistence curves. From Vink and Horbach \cite{81}. \label{fig1}}
\end{center}
\end{figure}

In the AO model \cite{73,74} colloids are described as hard
spheres of radius $R_{\rm c}$, and hence the potential between two
colloidal particles at distance $r$ from each other is
\begin{equation} \label{eq1}
U_{\rm cc}(r)=\infty \; (r < 2 R_{\rm c}), \ U_{\rm cc} (r)=0 \; ({\rm else}) 
\quad .
\end{equation}
Similarly, polymers are described as soft spheres of radius $R_{\rm p}$.
Remembering that long polymer chains with $N$ subunits have a
radius $R_{\rm p} \propto N^\nu$ with $ \nu \approx 0.59$ in good
solvent conditions \cite{56} or $R_{\rm p} \propto N^{1/2}$ in Theta
solvents \cite{56}, the density $\rho_{\rm p}=N/R^3_p$ of monomers of a
chain inside its own volume is very small, and hence polymer coils
can interpenetrate each other with a free energy cost of a few
$k_BT$ (with $k_B$ the Boltzmann constant and $T$ the temperature) \cite{84}. 
In the AO model, this free energy cost is
neglected, and the polymers are treated like particles in an ideal
gas, $U_{\rm pp}(r)=0$ irrespective of distance. But, of course,
polymers cannot penetrate into the colloidal particles, and hence
\begin{equation} \label{eq2}
U_{\rm pc} (r) = \infty \; (r < R_{\rm c} + R_{\rm p}) , \, \, U_{\rm pc} (r)=0 \; ({\rm
else}) .
\end{equation}
As is well-known \cite{40}, the polymers cause an (entropic)
depletion attraction between the colloidal particles, and as a
result, an entropy-driven phase separation occurs, if the volume
fractions $\eta_{\rm c}$, $\eta_{\rm p}$ of colloids and polymers are
sufficiently high (Fig.~\ref{fig1}). Here $\eta_{\rm c}$, $\eta_{\rm p}$ are
defined in terms of the volume $V$ of the system and the numbers
of colloids and polymer, $N_{\rm c}$ and $N_{\rm p}$, respectively, by
\begin{equation} \label{eq3}
\eta_{\rm c}= \frac{4 \pi} {3} R^3_{\rm c} N_{\rm c} / V \quad , 
\quad  n_p=\frac{4 \pi}{3} R^3_{\rm p} N_{\rm p}/V \, ,
\end{equation}
and $R_{\rm c}=1$ will henceforth be chosen as unit of length. Since
both $\eta_{\rm c}$, $\eta_{\rm p}$ are densities of extensive thermodynamic
variables, it is useful to carry out a Legendre transform to an
intensive thermodynamic variable, where the chemical potential
$\mu_{\rm p}$ of the polymers [or their fugacity $z_{\rm p}=\exp
(\mu_{\rm p}/k_BT)$] is used. It is customary to use instead of $\mu_{\rm p}$ or
$z_{\rm p}$ the so-called ``polymer reservoir  packing fraction''
$\eta^{\rm r}_{\rm p}$,
\begin{equation} \label{eq4}
\eta^{\rm r}_{\rm p} = z_{\rm p} \Big(\frac{4 \pi}{3} \Big) R^3_{\rm p} \, \;
\end{equation}
Eq.~(\ref{eq4}) would be just the volume fraction of polymers in
the absence of any colloids, since such a system simply is an
ideal gas of polymers.

It is clear that the model defined by
Eqs.~(\ref{eq1}),~(\ref{eq2}) is a drastic simplification of
reality, but in qualitative respects it is remarkably accurate
\cite{40}. While various more realistic extensions of the AO model
have been considered \cite{79,84,85,86,87,88,89,90,91,92}, and
sometimes better agreement with experiments \cite{44,93} is
obtained, we disregard such extensions here because in practice
there are many additional effects (such as charges on the
colloidal particles \cite{94,95}, adsorption of polymers on the
colloids \cite{96}, etc.) that make a quantitative comparison with
experiment elusive.

In early simulation work on the AO model \cite{77,78} a wider
range of volume fractions $\eta_{\rm c}$, $n_{\rm p}$ (and a much wider range
of $\eta^{\rm r}_{\rm p}$) was studied, but only a much more limited accuracy
than shown in Fig.~\ref{fig1} was obtained. On the basis of this
work \cite{77,78}, it was concluded that the agreement between
simulations and the mean-field theory of Lekkerkerker et
al.~\cite{75} is excellent. Fig.~\ref{fig1} demonstrates, however,
that the relative deviation between the actual value for
$\eta^{\rm r}_{\rm p}$ at the critical point, $\eta^{\rm r}_{\rm p, cr} =0.766 \pm
0.002$, deviates from its mean-field prediction \cite{75} by about
30\%. This deviation, in fact, is relatively larger than
corresponding deviations between mean field theory and accurate
simulation results for lattice gas models \cite{97}, Lennard-Jones
fluids \cite{98}, etc. In retrospect, this large deviation between
mean-field theory \cite{75} and accurate simulation results for
colloid-polymer mixtures \cite{80,81,82} is not surprising, since
on the length scale of a colloidal particle the depletion
attraction has a very short range, and the large absolute size of
colloidal particles in this context is not relevant: it would be
wrong to infer that colloids should behave mean-field like.

Being interested in the changes in phase behavior due to
confinement between walls that are a distance $D \gg R_{\rm c} (=1)$
apart, relatively small changes must be expected, of course. For
an analysis of these changes, and in particular for a study how
bulk behavior in the limit $D \rightarrow \infty$ is approached, a
very good accuracy of the simulation data is absolutely crucial.
Thus, it is worthwhile to briefly recall how results such as those shown
in Fig.~\ref{fig1} can be obtained, since the methods for the
study of the confined systems \cite{61,62,63} are closely related
to those used in the bulk \cite{80,81,82}.

We start this recollection by emphasizing that for studying
liquid-vapor type phase equilibria the grand-canonical $(\mu VT)$
ensemble of statistical mechanics is the best choice \cite{98},
since it avoids problems due to slow relaxation of liquid-vapor
interfaces that hamper the use of the canonical ensemble
\cite{99,100}. Also near the critical point the problem of
critical slowing down \cite{101} is somewhat less severe in the
grand-canonical ensemble \cite{98,99,100}, and the inevitable
finite size effects are relatively easy to handle by finite size
scaling methods \cite{28,29,30,102,103}, unlike the popular Gibbs
ensemble \cite{104,105}. So the task of the simulation is to vary
the chemical potential $\mu$ of the colloids at fixed $\eta^{\rm r}_{\rm p}$
(as the phase diagram, Fig.~\ref{fig1}, suggests, $\eta^{\rm r}_{\rm p}$ is
analogous to inverse temperature for ordinary vapor-liquid type
transitions \cite{98}, where vapor-liquid phase separation is
driven by enthalpic rather than entropic forces. Of course, in
thermal equilibrium the average colloid fraction $\langle \eta_{\rm c}
\rangle$, which is the variable thermodynamically conjugate to
$\mu$ (apart from a normalization factor, see Eq.~\ref{eq3}),
increases monotonously with $\mu$ even when the two-phase
coexistence region is crossed, and in the $\langle \eta_{\rm c}
\rangle$ vs.~$\mu$ curve hence no singularity shows up for any
finite linear dimension $L$: only in the thermodynamic limit
(where $L \rightarrow \infty$) this ``isotherm'' develops at
$\mu=\mu_{\rm coex}$ a perpendicular part, where $\langle \eta_{\rm c}
\rangle$ jumps discontinuously from $\eta_{\rm c}^{\rm V}$ (vapor) to
$\eta^{\rm L}_{\rm c}$ (liquid). However, nevertheless phase coexistence is
easily recognizable also in a finite volume simulation, when the
colloid volume fraction distribution $P(\eta_{\rm c})$ is sampled
\cite{98,99,100}. In the regime $\eta^{\rm V}_{\rm c} \leq \langle \eta \rangle
\leq \eta^{\rm L}_{\rm c}$, $P(\eta_{\rm c})$ has a double peak structure, and for
$\mu=\mu_{\rm coex}$ both peaks have equal weight 
(``equal area rule'' \cite{106,107}).

\begin{figure}
\begin{center}
\includegraphics[width=0.45\textwidth]{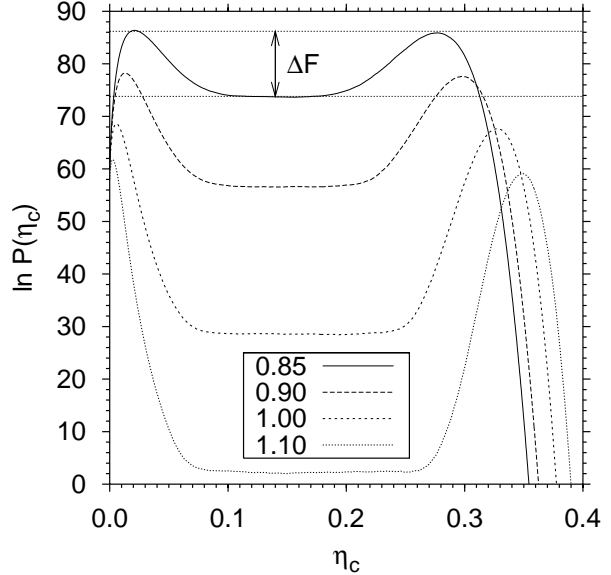}
\caption{Logarithm of the probability $P(\eta_{\rm c})$ of observing a
colloid packing fraction $\eta_{\rm c}$ for an AO mixture with $q=0.8$
at coexistence for several values of $\eta^{\rm r}_{\rm p}$ as indicated. The
simulations were performed in a box with linear dimensions
$L_x=L_y=16.7$ and $L_z=33.4$ using periodic boundary conditions.
Note that the distributions are not normalized, and that for
$\eta$ near $\eta_{\rm d}=(\eta^{\rm V}_{\rm c} + \eta^{\rm L}_{\rm c})/2$ the distribution is
essentially flat, almost independent from $\eta_{\rm c}$, so that a free
energy barrier $\Delta F$ is well-defined. From Vink and Horbach
\cite{81}. \label{fig2}}
\end{center}
\end{figure}

In order to carry out this program, two obstacles need to be
overcome: (i) in order to sample the relative weights of the two
peaks of $P(\eta_{\rm c})$, the peak near $\eta^{\rm V}_{\rm c}$ representing the
vapor-like phase of the colloid-polymer mixture and the peak near
$\eta^{\rm L}_{\rm c}$, the liquid-like phase, the system needs to cross many
times a region of very low probability near $\eta_{\rm d}=(\eta^{\rm V}_{\rm c} +
\eta^{\rm L}_{\rm c})/2$. This problem, however, can be very efficiently
solved by successive umbrella sampling \cite{108}. Fig.~\ref{fig2}
shows, as a typical example, distributions $P(\eta_{\rm c})$ that span
almost 30 decades. (ii) The second obstacle is the fact that the
polymer volume fraction, in the polymer-rich phase, can be very
high (exceeding unity, since the polymers are allowed to overlap
with no energy cost). Insertion of a colloid particle at a
randomly chosen position, which is one of the Monte Carlo (MC)
moves that one needs to carry out in grand-canonical Monte Carlo
simulations, almost always will be rejected: so a naive
implementation of a grand-canonical MC simulation for unfavorable
parameters is bound to fail utterly. However, this problem also
could be overcome, by the invention of a composite MC move, where
in a spherical region with some properly chosen radius $r_{\rm c}$ a
randomly selected chosen number $n_{\rm r}$ of polymers is taken out and
only then insertion of a colloid is attempted (the reverse move
also exists and is constructed such that the detailed balance
principle \cite{99,100} is fulfilled) \cite{81,82}.

\begin{figure}
\begin{center}
\includegraphics[width=0.45\textwidth]{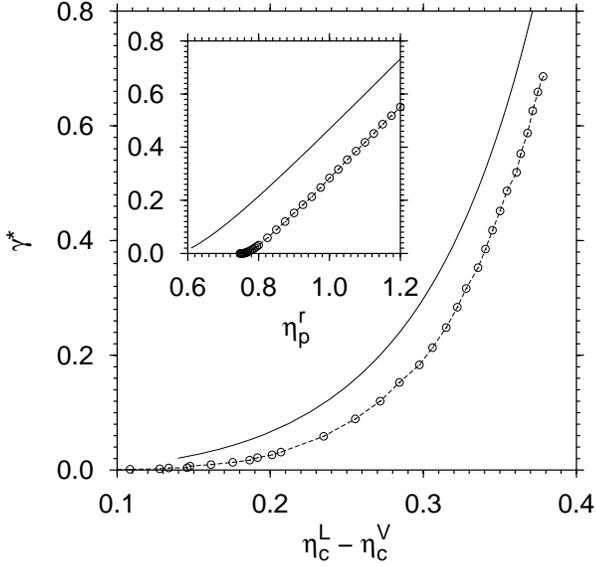}
\caption{Reduced interfacial tension 
$\gamma^*=4R^2_{\rm c} \gamma_{\rm LV}/k_BT$ for an AO mixture 
with $q=0.8$ as a function of
the difference in packing fraction between the coexisting phase,
$\eta^{\rm L}_{\rm c}-\eta^{\rm V}_{\rm c}$. The inset shows 
$\gamma^*$ vs.~$\eta^{\rm r}_{\rm p}$.
Open circles are the simulation results, while the full curves are
density functional theory predictions. From Vink and Horbach
\cite{81} \label{fig3}}
\end{center}
\end{figure}

We now return to the observation of Fig.~\ref{fig2}, that a high
free energy barrier $\Delta F$ (choosing units where $k_BT=1$)
exists, which is independent of $\eta_{\rm c}$ in a broad regime of
$\eta_{\rm c}$ around the composition of the rectilinear diameter
$\eta_{\rm d}$. The interpretation of this fact is that the system in
this region is in a state with two domains, separated by two
domain walls, oriented perpendicular to the $z$-direction, and
connected into itself by the periodic boundary conditions. This is
also confirmed by direct inspection of the configurations of the
system. Hence \cite{109}
\begin{equation} \label{eq5}
\Delta F = 2 L^2 \gamma_{\rm LV}, \quad L \rightarrow \infty \, ,
\end{equation}
where $\gamma_{\rm LV}$ is the interfacial tension between liquid- and
gas-like phases, and $L^2$ the interfacial area. Thus, estimating
$\Delta F$ for a series of cross-sectional areas $L^2$ of the
simulation box and extrapolating the result for $\gamma_{\rm LV}$ to
the thermodynamic limit has become a standard method for the MC
estimation of interfacial free energies \cite{99,100}.
Fig.~\ref{fig3} shows typical results for the reduced interfacial
tension plotted vs.~the order parameter $\eta^{\rm L}_{\rm c} - \eta^{\rm V}_{\rm c}$ and
compares them to density functional theory predictions \cite{110}.
These simulation results for $\gamma_{\rm LV}$ are also consistent
with a capillary wave analysis \cite{83}. Note that the
coexistence densities $\eta^{\rm V}_{\rm c}$, $\eta^{\rm L}_{\rm c}$ do approach the
predictions from mean field theory rather fast (Fig.~\ref{fig1}),
for $\eta^{\rm r}_{\rm p} \geq 1.0$ the differences are practically
invisible, however, no such convergence is seen for the
interfacial tension (Fig.~\ref{fig3}). The reason for the strong
discrepancies in Fig.~\ref{fig3} is not clear.

\begin{figure}
\begin{center}
\includegraphics[width=0.45\textwidth]{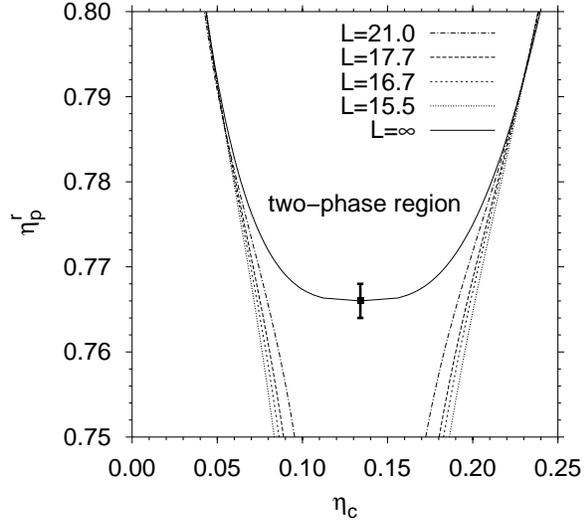}
\caption{Close-up of the phase diagram of the AO model for $q=0.8$
in the critical region, showing only the range $0.75 \leq\eta^{\rm r}_{\rm p}
\leq=0.80$ for the polymer reservoir packing fraction. Results for
four different choices of the linear dimension $L$ of $L \times L
\times L$ simulation boxes are shown as broken and dotted lines.
The extrapolation towards the thermodynamic limit, $L = \infty$,
also is included, and the estimate for the critical point
$\eta^{\rm r}_{\rm p,cr}=0.766\pm0.002$ is shown by the error bar. From Vink
et al.~\cite{82} \label{fig4}}
\end{center}
\end{figure}

We now comment on the treatment of finite size effects. If one
naively would take the values of $\eta_{\rm c}$ where $P(\eta_{\rm c)}$ has
its two peaks as estimates for $\eta^{\rm V}_{\rm c}$ and $\eta^{\rm L}_{\rm c}$ also in
the critical region, one obtains results as shown in
Fig.~\ref{fig4}: For $\eta_{\rm p}^{\rm r} \geq 0.79$ these estimates are
independent of the linear dimension of the simulation box, but for
$\eta^{\rm r}_{\rm p} < 0.79$ systematic finite size effects appear. E.g., for
$\eta^{\rm r}_{\rm p}=0.76$ the difference $\eta^{\rm L}_{\rm c} - \eta^{\rm V}_{\rm c}$ decreases
systematically with increasing $L$. While for $\eta^{\rm r}_{\rm p} >
\eta^{\rm r}_{\rm p,cr}$ this difference for $L \rightarrow \infty$
converges to a nonzero result, for $\eta^{\rm r}_{\rm p} \leq \eta^r_{\rm p,cr}$
it ultimately vanishes. While a naive inspection of
Fig.~\ref{fig4} does not allow to estimate $\eta^{\rm r}_{\rm p,cr}$, such
an estimate can be obtained reliably from finite size scaling
methods \cite{28,29,30, 87,88,89,90,100,102,103}. Choosing
$\mu=\mu_{\rm coex} (\eta^{\rm r}_{\rm p})$ from the equal area rule, as is
done in Figs.~\ref{fig1}, \ref{fig4}, we define an order parameter
$m$ as $m=\eta_{\rm c}- \langle \eta_{\rm c} \rangle$ and define moments
$\langle |m|^k \rangle$ ($k$ being integer) from the distribution
$P(\eta_{\rm c})$,
\begin{equation} \label{eq6}
\langle |m|^k \rangle = \int\limits^1_0  d \eta_{\rm c} \; |m|^k P(\eta_{\rm c})
\, .
\end{equation}
Defining then the fourth order cumulant $U_4$ as \cite{97,103}.
\begin{equation} \label{eq7}
U_4= \langle m^2 \rangle^2 / \langle m^4 \rangle
\end{equation}
we can invoke the result that $U_4$ tends towards unity for
$\eta^{\rm r}_{\rm p} > \eta^{\rm r}_{\rm r,cr}$ as $L \rightarrow \infty$, while $U_4$
tends to 1/3 for $\eta^{\rm r}_{\rm p} < \eta^{\rm r}_{\rm p, cr}$, since ultimately the
distribution $P(\eta_{\rm c})$ in the one-phase region must become a
single Gaussian centered at $\langle \eta_{\rm c} \rangle$ \cite{103}.
For $\eta^{\rm r}_{\rm p} =\eta_{\rm p, cr}$,however, $U_4$ tends to a nontrivial
but universal value $U_4^*$ ($U^*_4 \approx0.629$ in $d=3$
dimensions while $U^*_4 \approx 0.856$ in $d=2$ dimensions
\cite{111}). Consequently, plotting $U_4$ versus $\eta^{\rm r}_{\rm p}$ for
different $L$ one expects a family of curves that intersect at
$\eta^{\rm r}_{\rm p} = \eta_{\rm p,cr}$ in a common intersection point, if $L$ is
large enough so that corrections to finite size scaling are
negligible, and using this method (or an analogous reasoning
\cite{112} for the moment ratio $M= \langle m^2  \rangle / \langle
|m| \rangle^2 $ \cite{80,81} which should yield a universal
intersection in $d=3$ at $M^* =1.239$ \cite{113}) one finds the
estimate of $\eta^{\rm r}_{\rm p,cr}$ included in
Figs.~\ref{fig1},~\ref{fig4}.

\begin{figure}
\begin{center}
\includegraphics[width=0.45\textwidth]{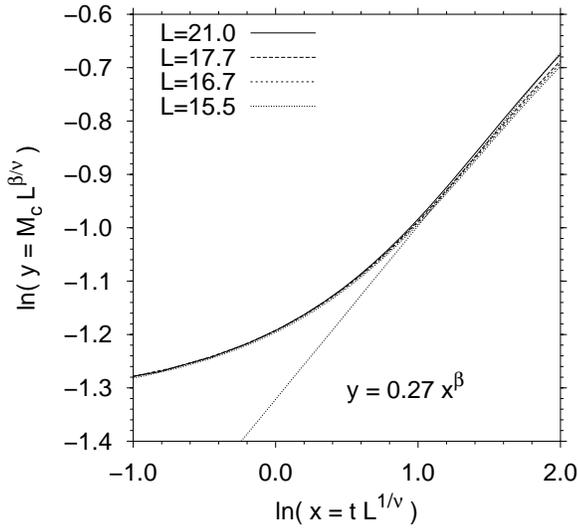}
\caption{Scaling plot for the order parameter $M_{\rm c} = \langle |m| \rangle$
of the colloid-polymer mixture in the bulk, for the AO model with
$q=0.8$, using only data for $t>0$. The quantity $L^{\beta/\nu}
M_{\rm c}$ is plotted vs.~$t L^{1/\nu}$, choosing logarithmic scales.
The straight line is a power law $y=0.27x^\beta$, with
$\beta=0.326$. From Vink et al.~\cite{83}. \label{fig5}}
\end{center}
\end{figure}

A further consequence of finite size scaling
\cite{28,29,30,97,98,99,100,102,103} is the fact that the moments
$\langle |m|^k \rangle$ are homogeneous functions of the two
variables $L$ and $t=\eta^{\rm r}_{\rm p} / \eta^{\rm r}_{\rm p,cr}-1$,
\begin{equation} \label{eq8}
\langle |m|^k \rangle =L^{-k \beta/\nu} \mathcal{M}_k (x), \quad x
= tL^{1/\nu} \, ,
\end{equation}
where $\beta$ and $\nu$ are the critical exponents of the order
parameter $M_c$ and correlation length $\xi$, respectively,
\begin{equation} \label{eq9}
M_{\rm c}=Bt^\beta , \quad \xi=  \hat{\xi} t^{-\nu} \; .
\end{equation}
In Eq.~(\ref{eq8}), $\mathcal{M}_k(x)$ is a scaling function, and $B$ and
$\hat{\xi}$ are critical amplitudes. For the universality class of
the $d=3$ Ising model \cite{114}, the exponents are
\cite{97,115,116}
\begin{equation} \label{eq10}
\beta \approx0.326, \quad \nu \approx0.630,
\end{equation}
which differ from the corresponding mean-field results \cite{114}
\begin{equation} \label{eq11}
\beta_{\rm MF}=1/2, \quad \nu_{\rm MF} =1/2 \quad .
\end{equation}
Taking the estimates for the exponents [Eq.~(\ref{eq10})] and 
using $\eta^{\rm r}_{\rm p,cr} =0.765$ we can replot the data 
of Fig.~\ref{fig4} in scaled form
(Fig.~\ref{fig5}), and indeed the data collapse rather well on a
master curve, as implied by Eq.~(\ref{eq8}). If we use
Eq.~(\ref{eq11}) instead, no such data collapsing is obtained.
This result shows that finite size scaling holds, and the AO model
also falls in the $d=3$ Ising universality class, as one might
have expected. Moreover, the straight line behavior seen on the
log-log plot for large $x$ not only implies that the data indeed
are compatible with the power law, $M_{\rm c}=Bt^\beta$, but
also the critical amplitude $B$ can be estimated with reasonable
accuracy, $B=0.27 \pm 0.02$ \cite{83}. This power law actually has
been included in Fig.~\ref{fig4} for $\eta^{\rm r}_{\rm p}$ near
$\eta^{\rm r}_{\rm p,cr}$. It results from Eq.~(\ref{eq8}) as the asymptotic
behavior for $L \rightarrow \infty$.

As is evident from the insert of Fig.~\ref{fig1}, the fluctuations
that are ignored by mean-field theory \cite{75} have two effects:
one effect is that the critical point $\eta^{\rm r}_{\rm p,cr}$ is shifted
upward (the compatibility of the colloid-polymer mixture is
enhanced), and the coexistence curve is flattened near the
critical point [according to mean-field theory, Eq.~(\ref{eq11}),
it is a simple quadratic parabola].

A similar discussion can be given for the interfacial tension,
$\gamma_{\rm LV}$ (Fig.~\ref{fig3}), which is found to vary as
\cite{83}
\begin{equation} \label{eq12}
\gamma_{\rm LV} =\hat{\gamma} \, t^\mu, \quad \mu=1.26,
\quad \hat{\gamma} \approx 0.26 \pm 0.02,
\end{equation}
while mean-field theory would imply $\mu =3/2$ \cite{114}. Vink et
al. \cite{83} have also analyzed the critical behavior of
susceptibilities at both sides of the transition and studied the
rectilinear diameter $\eta_{\rm d}$, as well as a few critical amplitude
ratios. All these analyses did confirm the Ising character of the
transition, indicating that the Ising critical region in fact is remarkably wide.
Mean-field theory \cite{75} is only reliable very far away from
criticality.

\section{Confinement by Symmetric Walls: Evidence for
Capillary-Condensation-Like Behavior}
In this section we consider colloid-polymer mixtures in a $L
\times L \times D$ geometry, where confinement is effected by two
identical walls a distance $D$ apart. In the simulations, we apply
periodic boundary condition in the $x$ and $y$-directions parallel
to the walls, and again the strategy will be to carry out an
extrapolation to the thermodynamic limit via a finite size scaling
analysis.

If one simply uses hard walls for both colloids and polymers, as
done in \cite{61}, one encounters a very pronounced depletion
attraction between the colloids and the walls, giving rise to a
very strong ``capillary condensation''-like shift \cite{10,11} of
the coexistence chemical potential $\mu_{\rm coex} (\eta^{\rm r}_{\rm p})$ of
the colloids. It is hence convenient to apply in addition a
square-well repulsive potential
\begin{equation} \label{eq13}
U_{\rm cw} (h)= \varepsilon, \; R_{\rm c} <h < 2 R_{\rm c}, 
\; U_{\rm cw} (h > 2 R_{\rm c}) =0,
\end{equation}
with $h$ the distance of a colloidal particle from the (closest)
wall. Of course, $U_{\rm cw} (h \leq R_{\rm c})=\infty$ and $U_{\rm pw} (h \leq
R_{\rm p})= \infty$, since neither colloids nor polymers are allowed to
penetrate into the wall.

If one considers very large $\varepsilon$, colloids are excluded from the
close vicinity of the walls, and an effective attraction of the
polymers to the walls would result. As a consequence, ``capillary
evaporation'' is expected rather than ``capillary condensation''
(i.e., close to phase coexistence in the bulk the capillary
prefers the vapor-like phase rather than the liquid-like phase of
the colloid-polymer mixture). Schmidt et al. \cite{59} presented a
(somewhat qualitative) evidence for this phenomenon.

Since the finite size thickness $D$ limits growth of the
correlation length $\xi$ of volume fraction fluctuations near the
critical point in the $z$-direction perpendicular to the confining
wall, a divergence of $\xi$ as described in Eq.~(\ref{eq9}) is
only possible along the $x$- and $y$-directions parallel to the
walls. Therefore, the phase transition which can take place is a
phase separation in lateral directions ($x,y$) only, between
colloid-rich and colloid pure phases. As a consequence, ultimately
this transition should belong to the universality class of the two-dimensional Ising model
\cite{114}, and the critical exponents are
\begin{equation} \label{eq14}
\beta=1/8, \quad \nu=1, \quad \mu=1,
\end{equation}
instead of those quoted in Eqs.~(\ref{eq10}),~(\ref{eq12}).
However, this two-dimensional critical behavior prevails only when
$\xi$ has grown to a size much larger than $D$: if $\xi \ll D$ the
behavior is still close to three-dimensional, and when $\xi$ and
$D$ are of the same order a gradual crossover between the two
types of critical behavior occurs.

\begin{figure}
\begin{center}
\includegraphics[width=0.45\textwidth]{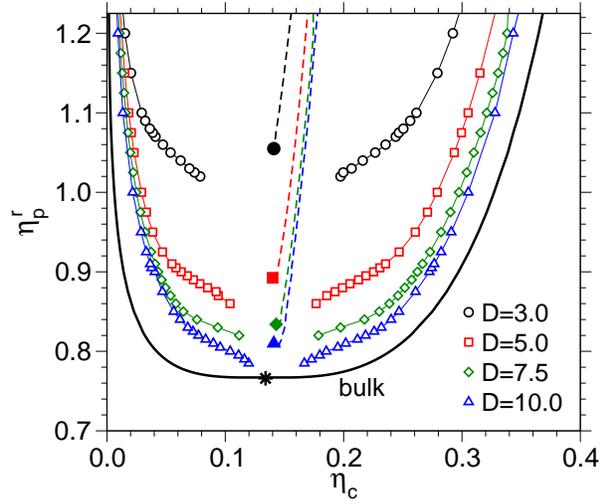}
\caption{Coexistence curves of the AO model with $q=0.8$ in the
bulk (full curve without data points, representing an
extrapolation to the thermodynamic limit) and for thin films
confined by symmetric walls, choosing Eq.~(\ref{eq13}) with
$\varepsilon=0.5$ as a wall potential. Open circles denote data
for $D=3$, $L=18$ (all lengths here are measured in units of the
colloid diameter, $\sigma_{\rm c}=2R_{\rm c}$). Open squares denote data for
$D=5$, $L=20$; open diamonds are for $D=7.5$, $L=30$ , and open
triangles for $D=10$, $L=30$. The broken curves denote the
corresponding coexistence diameters, $\eta_{\rm d}= (\eta^{\rm L}_{\rm c} +
\eta^{\rm V}_{\rm c})/2$ and the full symbols denote estimates of the
corresponding critical points for the various choices of $D$. From
Vink et al.~\cite{62}. \label{fig6}}
\end{center}
\end{figure}

\begin{figure}
\begin{center}
\includegraphics[width=0.45\textwidth]{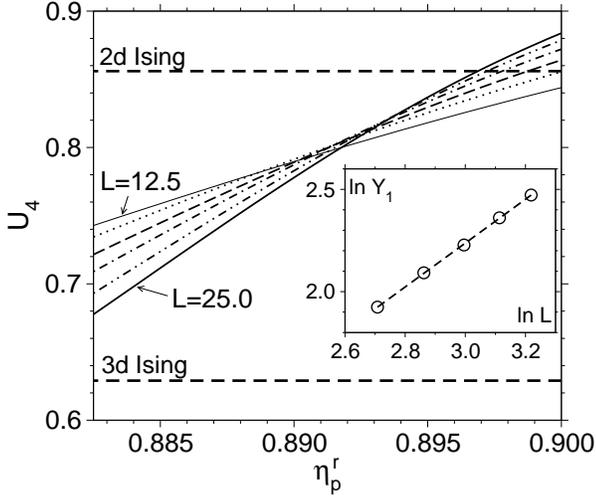}
\caption{Moment ratio $U_4$ for a film of thickness $D=5$ plotted
versus the polymer reservoir packing fraction $\eta^{\rm r}_{\rm p}$ and
several choices of the lateral linear dimension, $L=12.5$, 15,
17.5, 20, 22.5, and 25. The intersection points allow to estimate
the critical point as $\eta^{\rm r}_{\rm p,cr}=0.892 \pm 0.002$. The inset
shows a log-log plot of the slope $Y_1= dU_4/d \eta^{\rm r}_{\rm p}$ at
$\eta^{\rm r}_{\rm p,cr}$ versus $L$, testing the prediction that $Y_1
\propto L^{1/\nu}$. The straight line in the inset corresponds to
an (effective) exponent $\nu_{\rm eff} =0.933$. The broken
horizontal straight lines in the main part indicates the
theoretical values for the universal value $U^*$ of $U_4$ at the
intersection point, for both the $d=2$ and the $d=3$ Ising model,
respectively. From Vink et al.~\cite{62}. \label{fig7}}
\end{center}
\end{figure}

These crossover phenomena make the analysis of the simulations
somewhat more difficult. For any finite value of $L$ the ``raw
data'' estimates for $\eta^{\rm V}_{\rm c}$, $\eta^{\rm L}_{\rm c}$ are qualitatively
similar, irrespective of $D$ (Fig.~\ref{fig6}). Again pronounced
``finite size tails'' occur for these estimates in the vicinity of
$\eta^{\rm r}_{\rm p,cr}$, i.e., for any finite $L$ one finds that $\eta^{\rm V}_{\rm c}$ 
and $\eta^{\rm L}_{\rm c}$ as estimated from the peak positions of
$P(\eta_{\rm c})$ fail to merge at $\eta^{\rm r}_{\rm p, cr}$, but rather continue
further into the one-phase region, as in the bulk
(Fig.~\ref{fig4}). When one then plots $U_4$ vs.~$\eta^{\rm r}_{\rm p}$ for
different choices of $L$, searching for a universal intersection
point, one rather finds that the intersection points are somewhat
scattered over a region of values for $\eta^{\rm r}_{\rm p}$
(Fig.~\ref{fig7}). In addition, this intersection does occur
neither at the theoretical value for $U^*$ for the $d=2$
universality class nor at the $U^*$ for $d=3$, but rather
somewhere in between. These findings are a consequence of the
gradual crossover in critical behavior alluded to above. While for
$D=3$ both $\nu$ and $U^*$ are rather close to the theoretical
$d=2$ values, for $D=10$ both $\nu$ and $U^*$ are about half way
between the $d=2$ and $d=3$ values. However, these numerical
results do not have any fundamental significance; they only mean
that the larger $D$ the closer $\eta^{\rm r}_{\rm p,cr}$ needs to be
approached, to be in the region where ultimately $\xi \gg D$ and
hence the correct asymptotic critical behavior (which is always
two-dimensional, for any finite value of $D$) can be seen.

\begin{figure}
\begin{center}
\includegraphics[width=0.45\textwidth]{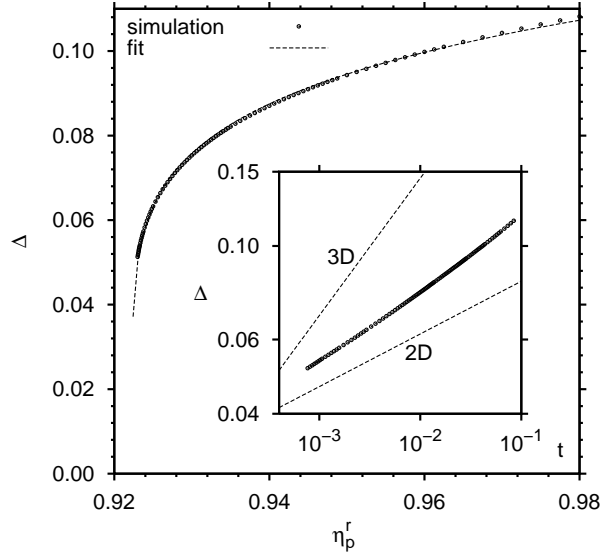}
\caption{Order parameter $\Delta (=M_{\rm c})$ of the confined AO model
with $q=0.8$ plotted vs.~$\eta^{\rm r}_{\rm p}$, choosing $D=5$ and
$\varepsilon=0$. The curve through the simulation data in the main
frame is a fit to $\Delta =B_{\rm eff} t^{\beta_{\rm eff}}$,
choosing $\eta^{\rm r}_{\rm p,cr} =0.9223$. The inset shows the same data as
a function of the relative distance from the critical point,
$t=\eta^{\rm r}_{\rm p}/\eta^{\rm r}_{\rm p,cr}-1$, on double logarithmic scales; broken
straight lines illustrate the two-dimensional and three
dimensional Ising exponents. From Vink et al.~\cite{61}.
\label{fig8}}
\end{center}
\end{figure}

For the case $D=5$, $\varepsilon=0$ a very careful analysis has
been performed \cite{61}, applying a novel variant of finite
scaling which does not imply any bias on the type of critical
exponents \cite{117,118,119}. Fig.~\ref{fig8} shows that the
resulting order parameter can be fitted over some range indeed by
an effective exponent $\beta_{\rm eff}=0.17$, which is in between
the $d=2$ and $d=3$ values (0.125 $< \beta _{\rm eff} < 0.326)$,
but a correct interpretation of this finding is that a log-log
plot of the order parameter vs.~$t$ exhibits a slight curvature,
spread out over several decades. Only for $t \rightarrow 0$ can
the $d=2$ value ($\beta =0.125)$ be expected to be seen; for
larger $t$ the slope $\beta_{\rm eff}$ on the log-log plot
increases systematically (but it does not reach the $d=3$ value,
since for $t \geq 0.1$ noncritical saturation effects come into play).

\begin{figure}
\begin{center}
\includegraphics[width=0.485\textwidth]{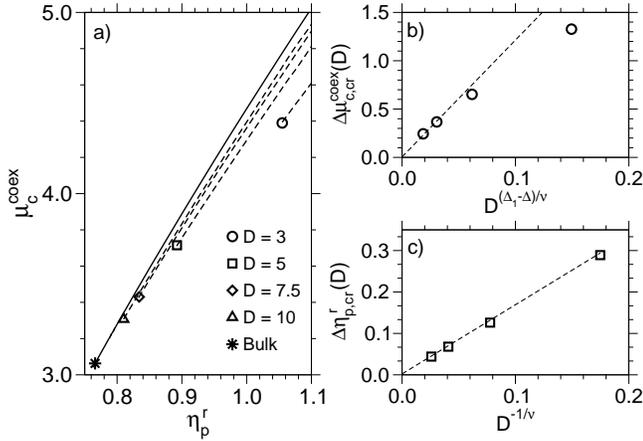}
\caption{a) Coexistence curves of the AO model with $q=0.8$ in
the grand-canonical representation where the chemical potential
$\mu^{\rm coex}$ of the colloids at the coexistence curve is
plotted vs.~the polymer reservoir packing fraction $\eta^{\rm r}_{\rm p}$. The
bulk result $(D=\infty)$ is shown as a full curve, while the
broken curves show the coexistence curves for confined films for
several thicknesses $D$, as indicated. The symbols mark the
corresponding critical points. b) Shift of the critical
coexistence colloid chemical potential plotted vs.~$D^{(\Delta_1-\Delta)/\nu}$. 
Equation (\ref{eq16}) implies a straight
line, as indicated by the dashed curve. c) Shift of the critical
polymer reservoir packing fraction, $\Delta \eta^{\rm r}_{\rm p,cr}(D)$,
plotted vs.~$D^{-1/\nu}$. The dashed lines indicate that
Eq.~(\ref{eq15}) holds. From Vink et al.~\cite{62}.
\label{fig9}}
\end{center}
\end{figure}

While the critical behavior of thin films is controlled by the
$d=2$ critical exponents, a different answer results when one
considers the shift of the critical point relative to the bulk
\cite{33,34}: this shift is controlled by three-dimensional
exponents only, namely
\begin{eqnarray}
\Delta \eta^{\rm r}_{\rm p,cr} (D) & \equiv & 
\eta^{\rm r}_{\rm p, cr}(D) - \eta^{\rm r}_{\rm p,cr}(\infty) \nonumber \\
& \propto & D^{-1/\nu}, \; D\to \infty \label{eq15} \\
\Delta \mu^{\rm coex}_{cr} (D) & \equiv & 
\mu^{\rm coex}_{\rm cr}(D) - \mu^{\rm coex}_{\rm cr}(\infty) \nonumber \\
& \propto & D^{(\Delta_1-\Delta)/ \nu}, \; D\to \infty. \label{eq16}
\end{eqnarray}
Here $\Delta \approx 1.56$ \cite{97,115,116} is the so-called
``gap exponent'' which characterizes the bulk equation of state
near criticality, and $\Delta_1 \approx 0.47$
\cite{120,121,122,123} its surface analog. Figure \ref{fig9} shows
that the AO model is compatible with these predictions, even
though only rather small film thicknesses were accessible to the
simulation (the largest film thickness included in Fig.~\ref{fig9}
is only for $D=10$ colloid diameters).

\begin{figure}
\begin{center}
\includegraphics[width=0.45\textwidth]{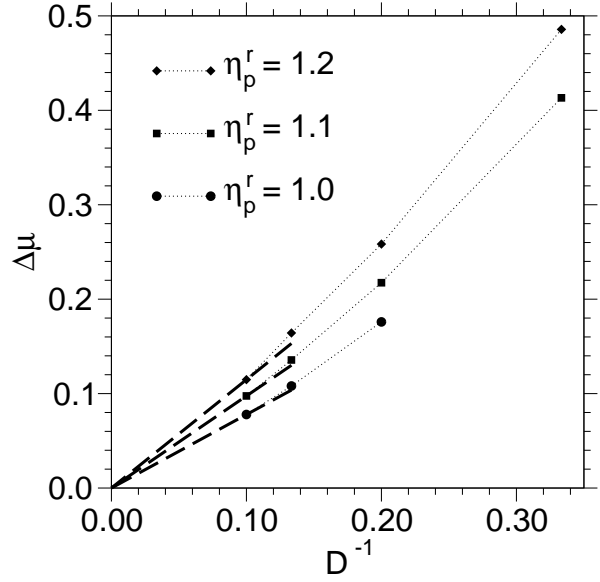}
\caption{Test of the Kelvin equation. The chemical potential
difference $\Delta \mu^{\rm coex} (D)$ \{Eq.~(\ref{eq17})\} is
plotted vs.~$D^{-1}$ for three values of $\eta^{\rm r}_{\rm p}$, chosen well
above the critical values $\eta^{\rm r}_{\rm p,cr}(D)$. Broken straight
lines show that the data are compatible with the Kelvin equation.
From Vink et al.~\cite{62}. \label{fig10}}
\end{center}
\end{figure}

Note that for $\eta^{\rm r}_{\rm p} > \eta^{\rm r}_{\rm p,cr} $ the asymptotic behavior
of the shift of the colloid chemical potential at phase
coexistence is not given by Eq.~(\ref{eq16}), but by the simpler
``Kelvin equation'' \cite{10,35}
\begin{eqnarray} \label{eq17}
\Delta \mu^{\rm coex} (D) & = & \mu^{\rm coex} (D) - \mu^{\rm coex}(\infty) \nonumber \\
& \propto & 1/ D, \,
D \rightarrow \infty, \, \eta^{\rm r}_{\rm p} > \eta^{\rm r}_{\rm p,cr}.
\end{eqnarray}
Fig.~\ref{fig10} shows that the data of Vink et al. \cite{62} are
compatible with this equation as expected.

We emphasize that in mean field theory one could not discuss the
crossover between two- and three-dimensional critical behavior,
since $\beta_{\rm MF} =1$ irrespective of dimensionality
\cite{114}, and also the mean-field predictions for the shift of
the critical point [Eqs.~(\ref{eq15}), (\ref{eq16})] would be
different from what was observed \cite{62}, since $1/\nu_{\rm
MF}=2$ instead of $1/\nu\approx1.59$, and ($\Delta^{\rm
MF}_1-\Delta^{\rm MF})/\nu_{\rm MF}=-2$ instead of
$(\Delta_1-\Delta)/\nu \approx-1.73$. However, mean-field theory
does reproduce the Kelvin equation, Eq.~(\ref{eq17}), and in any
case mean-field results for our confined films would be desirable.
We note that some mean-field results as well as Monte Carlo
results are available for $q=1$ \cite{58,59}; however, the
accuracy of these Monte Carlo data was too limited to allow for
a comprehensive test of theoretical predictions, as reviewed above,
and hence these studies \cite{58,59} are not discussed further
here.

\begin{figure}
\begin{center}
\includegraphics[width=0.485\textwidth]{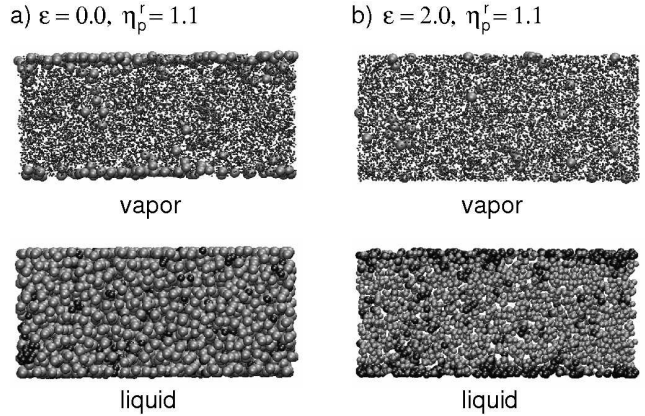}
\caption{Snapshot pictures of coexisting phases for the
colloid-polymer mixture with $q=0.8$, $D=10$, $\eta^{\rm r}_{\rm p}=1.1$, for
$\varepsilon=0$ (a) and $\varepsilon=2.0$ (b) Colloidal particles
are in green, polymers in blue (the size of the polymers is rescaled to allow
a clearer view). \label{fig11}}
\end{center}
\end{figure}

\begin{figure}
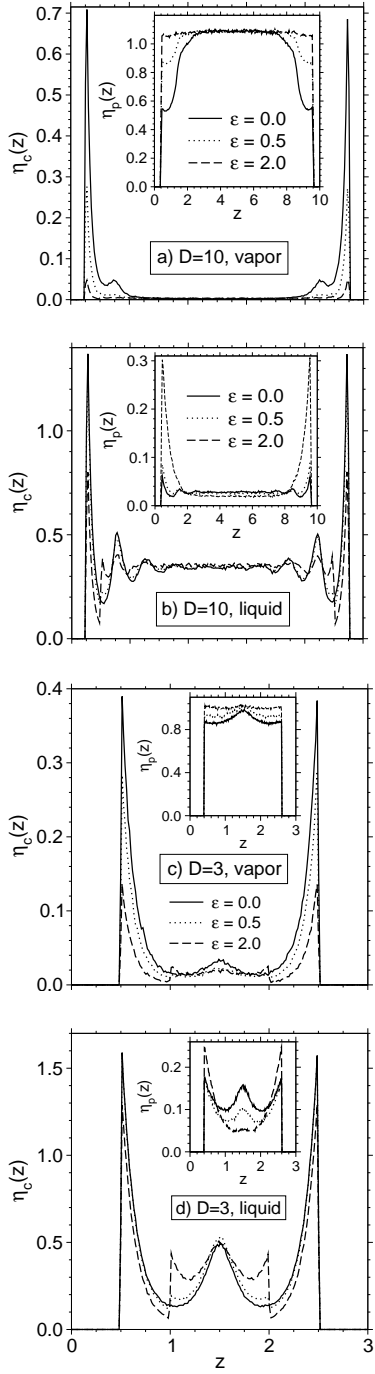

\begin{center}
\includegraphics[width=0.28\textwidth]{fig12a.eps}
\includegraphics[width=0.28\textwidth]{fig12b.eps}
\includegraphics[width=0.28\textwidth]{fig12c.eps}
\includegraphics[width=0.28\textwidth]{fig12d.eps}
\caption{Colloid density profiles obtained in thin films at
$\eta^{\rm r}_{\rm p}=1.1$, for two values of the film thickness $D$, and
several values of the colloid-wall parameter $\varepsilon$ as
indicated. Frames a) and b) show profiles obtained for $D=10$,
on the vapor and liquid branch of the coexistence curve,
respectively. Frames c) and d) show the corresponding profiles for
thickness $D=3$. Note the jumps in the colloid density at $z=0.5,
1.0,2.0,2.5$ caused through the jumps of the potential at $z=R_{\rm c}$ and
$z=2R_{\rm c}$, respectively. The insets represent density profiles of
the polymers. From Vink et al.~\cite{62}. \label{fig12}}
\end{center}
\end{figure}

We conclude this section by discussing the structure of the
coexisting phases in the thin film in more detail. Already
snapshot pictures [Fig.~\ref{fig11}] show that in the
$z$-direction the composition can be inhomogeneous. For
$\varepsilon=0$ colloidal particles are enriched at the walls in
the vapor-like phase, while for $\varepsilon=2$ polymers are
enriched at the walls in the liquid-like phase. However, it would
be wrong to consider these enrichment layers as wetting layers
\cite{24,25,26,27}: wetting layers are macroscopically thick, and
cannot occur in a thin film geometry \cite{11}. One should also
recall that wetting at the surface of a semi-infinite system
occurs at bulk coexistence, while coexistence in the thin film
deviates from bulk coexistence \{cf.
Eqs.~(\ref{eq16}), (\ref{eq17}), and
Figs.~\ref{fig9}a,~\ref{fig10}\}. Fig.~\ref{fig12} shows density
profiles across the thin film for several typical choices of
parameters. One recognizes that the colloid density in the liquid
-like phase near the walls shows a pronounced layering effect,
while the polymer density in the vapor-like phase lacks a
corresponding effect. This finding is expected, since layering is
a consequence of the repulsive interactions among the particles.
While for $D=10$ the films do reach homogeneous bulk-like states
in their center, for $D=3$ and $D=5$ (not shown) the behavior
stays inhomogeneous throughout the film.

\section{Confinement by Asymmetric Walls: Evidence for an
Interface Localization Transition}
\begin{figure}
\begin{center}
\includegraphics[width=0.485\textwidth]{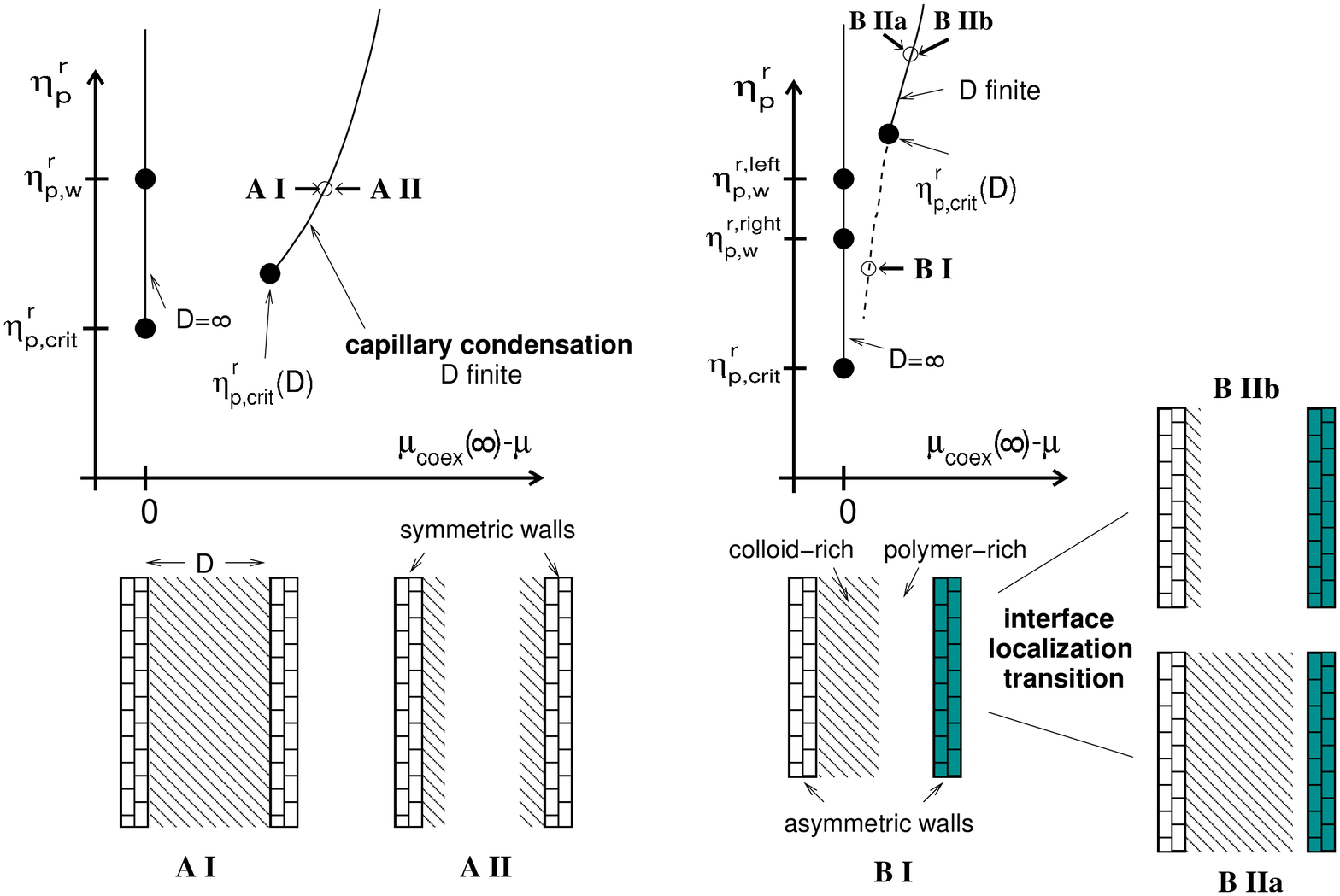}
\caption{Schematic phase diagram of a colloid-polymer mixture
confined between two parallel walls a distance $D$ apart, in the
grand-canonical ensemble where the polymer reservoir packing
fraction $\eta^{\rm r}_{\rm p}$ is used as ordinate and the difference between
the chemical potential of the colloids at bulk phase coexistence
$\mu_{\rm coex} (D=\infty)$ and the actual colloid chemical
potential is used as abscissa. Thus, phase coexistence in the bulk
occurs along a vertical straight line at $\mu_{\rm coex}(\infty)-\mu=0$. 
The left part shows the case of symmetric walls, the
right part asymmetric walls. States AI and AII coexist along the
capillary condensation transition line, states BIIa and BIIb coexist
along the interface localization line, while state BI exists along
the broken curve (which represents a line of rounded transitions).
In the limit $D \rightarrow \infty$, which corresponds to an
infinite system but bounded by walls both on the left and the
right side, wetting transitions occur, that are rounded off for
finite $D$. In the symmetric situation, the wetting transitions of
both walls coincide at $\eta^{\rm r}_{\rm p,w}$, while in the asymmetric
situation $\eta^{\rm r, right}_{\rm p,w} \neq \eta^{\rm r, left}_{\rm p,w}$. 
From De Virgiliis et al.~\cite{63}. \label{fig13}}
\end{center}
\end{figure}

By asymmetric walls one can realize a situation that one wall
attracts predominantly colloids and the other wall attracts
polymers. As discussed in the previous section, hard walls attract
colloids via a depletion mechanism; but coating the wall by a
polymer brush under semidilute conditions, one may cancel this
depletion attraction partially or completely, and also reach a
situation where polymers get attracted to the wall. This situation
is qualitatively modelled by a step potential of height
$\varepsilon$, acting on the colloids only \{Eq.~(\ref{eq13})\}.
An asymmetric situation occurs e.g. if the left wall is a hard
wall but on the right wall the additional potential described by
Eq.~(\ref{eq13}) acts, see Fig.~(\ref{fig13}). In drawing the
schematic phase diagrams, we have assumed that for a semi-infinite
system colloid-polymer mixtures exhibit complete wetting
\cite{11,24,25,26,27} over a wide range near the critical point,
namely for $\eta^{\rm r}_{\rm p,cr} \leq \eta^{\rm r}_{\rm p} \leq \eta^{\rm r}_{\rm p,w}$, while for
$\eta^{\rm r}_{\rm p} > \eta^{\rm r}_{\rm p,w}$ ``incomplete wetting'' (i.e., a nonzero
contact angle of a droplet) would occur. This assumption is
corroborated by density functional calculations \cite{124} and
Monte Carlo simulations \cite{105,125}. Since no ``prewetting
transition'' \cite{11,24,25,26,27} was found, the wetting
transition presumably is of second order, and this was assumed
drawing the phase diagrams of Fig.~\ref{fig13}, since this greatly
simplifies the theoretical analysis. For the case of symmetrical
mixtures of long flexible macromolecules, the influence of
prewetting phenomena on the phase diagram of thin confined films
has been thoroughly investigated
\cite{11,39,65,66,67,68,69,70,71,72}, and it has been shown that
typically a phase diagram with two critical points and a triple
point can be expected.

For asymmetric walls an interface localization transition may
occur, and this situation is explained qualitatively in the right
part of Fig.~\ref{fig13}. If the strength of the attraction of the
colloids of the left wall is of the same order as the strength of
the attraction of the polymers to the right wall, $\eta^{\rm r,
right}_{\rm p,w}$ and $\eta^{\rm r, left}_{\rm p,w}$ will be rather close
to each other and both exceed $\eta^{\rm r}_{\rm p,cr}$ distinctly. Then for
$\eta^{\rm r}_{\rm p} > \eta^{\rm r}_{\rm p,cr}$ the left wall will always be coated
with colloids, the right wall will always be coated with polymers.
In other words, we expect an interface between the colloid-rich
phase on the left and the polymer-rich phase on the right. When
$\mu$ is small enough \{i.e, $\mu_{\rm coex} (\infty) - \mu$ is
large enough\} most of the system is in the polymer-rich phase
(shown schematically as BIIb in Fig.~\ref{fig13}) but when $\mu$
increases a transition takes place to a state where most of the
film is in the colloid-rich phase (state BIIa). 
For $\eta^{\rm r}_{\rm p} > \eta^{\rm r}_{\rm p,cr}(D)$ this 
transition is a sharp (first-order) phase
transition, i.e.~the interface jumps from a state localized at the
left wall to a state localized at the right wall. 
For $\eta_{\rm p}^{\rm r} = \eta^{\rm r}_{\rm p,cr}(D)$ this 
transition is of first order, while for
$\eta^{\rm r}_{\rm p} <\eta^{\rm r}_{\rm p, cr}(D)$ the transition is a smooth gradual
transition (near the broken line in Fig.~\ref{fig13}). Note,
however, that this transition becomes sharper and sharper as $D$
increases, but a true phase transition appears only in a
discontinuous manner in the limit $D \rightarrow \infty$
\cite{11}: then the broken line in Fig.~\ref{fig13} coincides with
the line $\mu=\mu_{\rm coex} (\infty)$ ending at $\eta^{\rm r}_{\rm p,cr}$ ,
and $\eta^{\rm r}_{\rm p,cr}(D)$ does not converge to $\eta^{\rm r}_{\rm p,cr}$ but
rather we have $\eta^{\rm r}_{\rm p,cr} (D \rightarrow \infty)
= \eta^{\rm r, left}_{\rm p,w}$ (for the situation drawn in Fig.~\ref{fig13}).

\begin{figure}
\begin{center}
\includegraphics[width=0.49\textwidth]{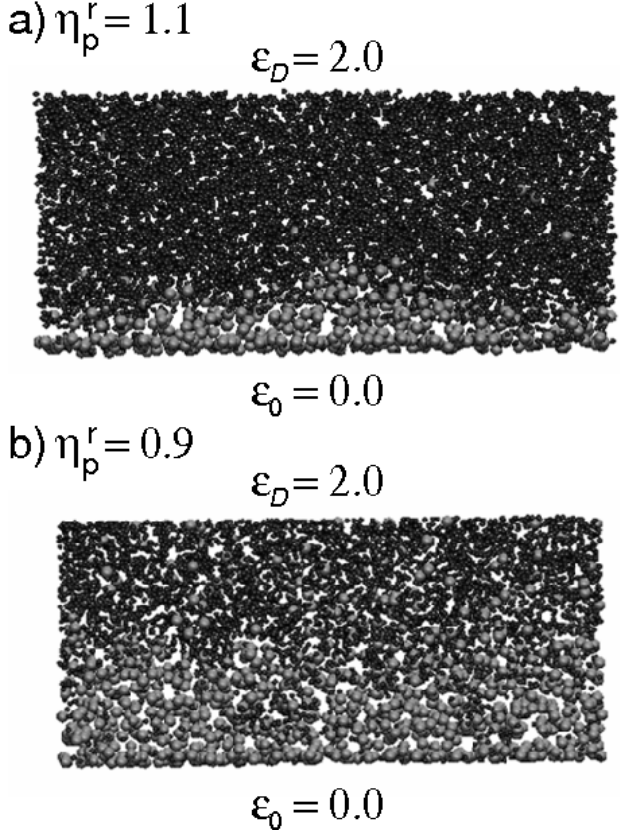}
\caption{a) Snapshot picture of the polymer-rich phase, for
$D=10$, $q=0.8$, $\eta^{\rm r}_{\rm p}=1.1$. The lower wall (at $z=0$) has a
wall potential parameter energy $\varepsilon_0=0$ and hence
attracts colloidal particles (shown in green), while the upper
wall (at $z=D$) has a wall potential energy parameter
$\varepsilon_D=2$, attracting polymers (shown in blue). b)
Snapshot picture of the same system as in a), but for
$\eta^{\rm r}_{\rm p}=0.9$ showing a state with a delocalized interface.
\label{fig14}}
\end{center}
\end{figure}

\begin{figure}
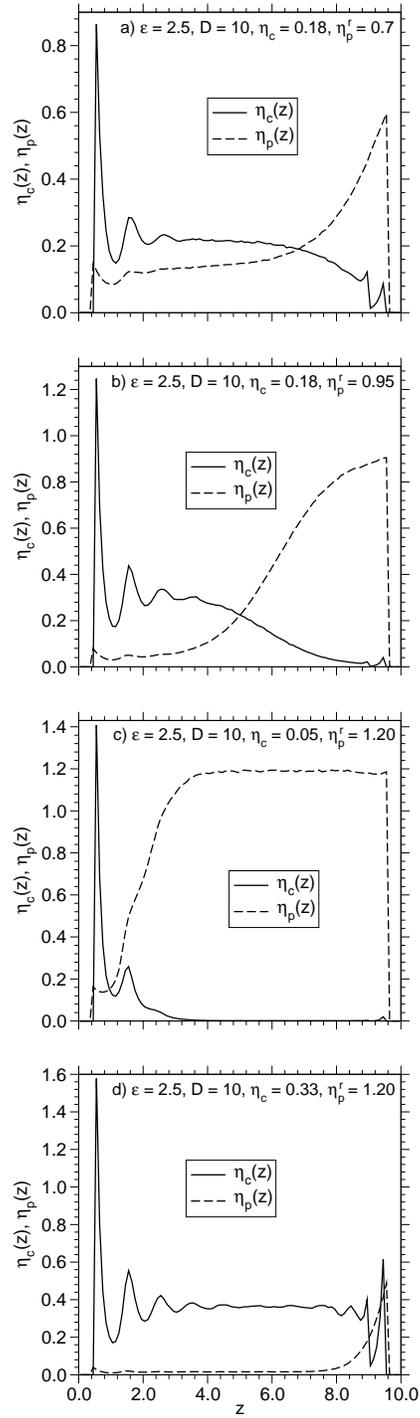

\begin{center}
\includegraphics[width=0.31\textwidth]{fig15a.eps}
\includegraphics[width=0.31\textwidth]{fig15b.eps}
\includegraphics[width=0.31\textwidth]{fig15c.eps}
\includegraphics[width=0.31\textwidth]{fig15d.eps}
\caption{Colloid concentration profiles $\eta_{\rm c}(z)$ and polymer
concentration profiles $\eta_{\rm p}(z)$ as a function of $z$ for a thin
film with asymmetric walls (hard wall at $z=0$, while for the
other wall at $z=D=10$ the potential $U_{\rm cw}(h)$ acts see
Eq.~(\ref{eq13}), with $\varepsilon=2.5$). Profiles were obtained
at $\eta_{\rm c}=0.18$, $\eta^{\rm r}_{\rm p} =0.70$ (a) , $\eta_{\rm c}=0.18$,
$\eta^{\rm r}_{\rm p}=0.95$ (b), $\eta_{\rm c}=0.05$, $\eta^{\rm r}_{\rm p} =1.20$ (c), and
$\eta_{\rm c}=0.33$, $\eta^{\rm r}_{\rm p}=1.20$ (d). For profiles (c) and (d), the
choices $\eta_{\rm c}=0.05$, $0.33$ roughly correspond to the two
branches of the coexistence curve, see Fig.~\ref{fig16}. From De
Virgiliis et al.~\cite{63}. \label{fig15}}
\end{center}
\end{figure}

For the states along the broken curve in Fig.~\ref{fig13} the
system is essentially inhomogeneous, there exists a thick domain
of colloid-rich phase in the left part of the film, and a thick
domain of polymer-rich phase in the right part, separated by a
``delocalized'' interface in the center of the film
\cite{11,36,37,38,39}. Snapshot pictures of the system indeed
readily confirm such a scenario (Fig.~\ref{fig14}), as well as the
density profiles across the thin film (Fig.~\ref{fig15}).
Fig.~\ref{fig15}a shows the profile for 
$\eta^{\rm r}_{\rm p}= 0.7 < \eta^{\rm r}_{\rm p,cr}$, 
i.e.~a state in the one phase region of the bulk.
One recognizes that the colloid concentration is enhanced near the
hard wall, as expected from the depletion attraction. Near the
other wall at $z=D=10$, the colloid concentration is somewhat
depressed, but the polymer concentration is clearly enhanced. But
in the center of the thin film both profiles are roughly constant,
as expected for bulk-like behavior. In fact, for large $D$ we
expect that the surface enhancement (or reduction, respectively)
decays with $z$ according to an exponential relation,
$\exp(-h/\xi)$, where $h$ is the distance from the closest wall
and $\xi$ the bulk correlation length \cite{120}.
Figure~\ref{fig15}b shows the profiles at $\eta^{\rm r}_{\rm p}=0.95$, which
exceeds the bulk critical value $\eta^{\rm r}_{\rm p,cr}$, but still is
smaller than the critical value $\eta^{\rm r}_{\rm p,cr}(D)$ of the confined
system. Now, the profiles are very different from those of
Fig.~\ref{fig15}a: phase separation in a colloid-rich and a
polymer-rich phase has occurred, with the interface position
(estimated from the inflection point of the polymer volume
fraction profile $\eta_{\rm p}(z)$, for instance) being located in
the center of the film. This situation corresponds to the snapshot
in Fig.~\ref{fig14}b. The interfacial profile resembles that of an
interface between bulk coexisting phases, broadened by capillary
waves \cite{68,83}. Finally, the cases $\eta^{\rm r}_{\rm p}=1.2$ 
(Fig.~\ref{fig15}c,d) refer to the two-phase region of the film. The
interface either is located near the hard wall, corresponding to
the polymer-rich phase of the film (Fig.~\ref{fig15}c: this case
corresponds to the snapshot shown in Fig.~\ref{fig14}a), or near
the wall that attracts the polymers (Fig.~\ref{fig15}d). Note that
along the transition line drawn schematically in Fig.~\ref{fig13}
(right part), there occurs lateral phase separation between the
states corresponding to these two types of profiles,
Fig.~\ref{fig15}c and Fig.~\ref{fig15}d, which hence can coexist
with each other in a thin film (and then are separated by an
interface running from the right wall towards the left wall).

\begin{figure}
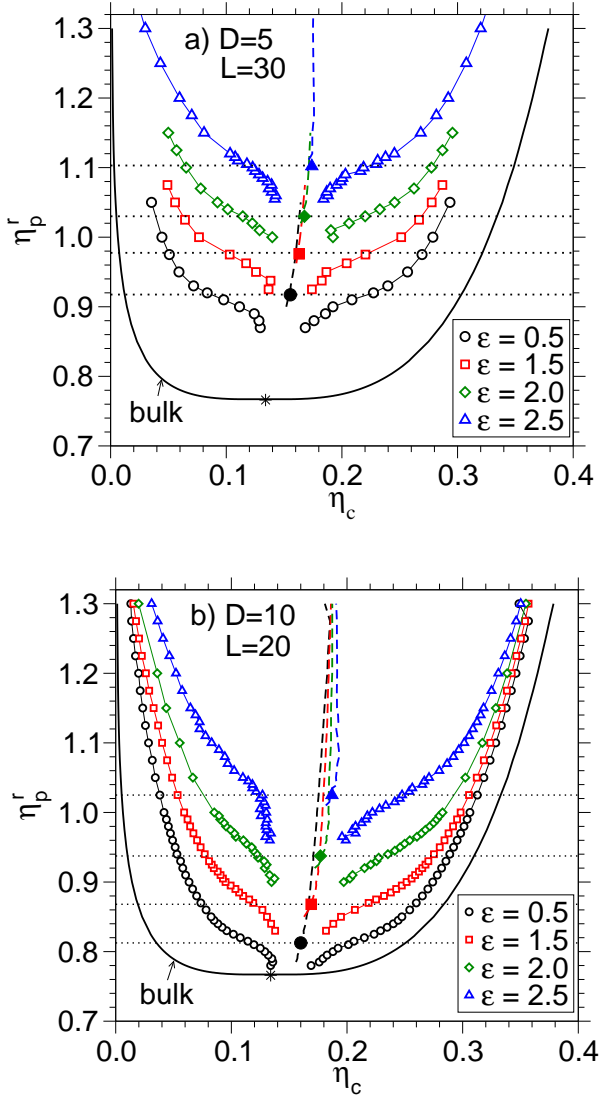

\begin{center}
\includegraphics[width=0.45\textwidth]{fig16a.eps}
\vspace*{0.7cm}

\includegraphics[width=0.45\textwidth]{fig16b.eps}
\caption{Coexistence curves for $D=5$, $L=30$ (a) and $D=10$,
$L=20$ (b), using four values of $\varepsilon$, as indicated. Also
the bulk coexistence curve is shown (full curves). Full symbols
mark critical points, the broken lines ending at these critical
points are the coexistence diameters. The dotted horizontal
straight lines mark the values of $\eta^{\rm r}_{\rm p, cr}(D)$. From De
Virgiliis et al.~\cite{63}. \label{fig16}}
\end{center}
\end{figure}

Figure \ref{fig16} shows the corresponding phase diagrams for two
film thicknesses, $D=5$ and $D=10$, varying the strength
$\varepsilon$ of the potential [Eq.~(\ref{eq13})] at the right
wall. One sees that with increasing $\varepsilon$ the critical points and
the whole coexistence curves are shifted upwards, to rather large
values of $\eta^{\rm r}_{\rm p}$. This shift is consistent with the
qualitative phase diagram of Fig.~\ref{fig13} (right part). Of
course, one must again recall that in Fig.~\ref{fig16} we show
``raw Monte Carlo data'' for one choice of $L$ only, and hence
pronounced finite size tails near $\eta^{\rm r}_{\rm p,cr} (D)$ are
apparent, as discussed for the case of capillary condensation
already (Fig.~\ref{fig6}). The critical points were again
estimated from the cumulant intersection method. Although strong
corrections to scaling are present, the conclusion can be drawn
\cite{63} that the critical behavior of the interface localization
belongs to the $d=2$ Ising universality class, as expected
\cite{36,37,38}.

\begin{figure}[ht]
\begin{center}
\includegraphics[width=0.45\textwidth]{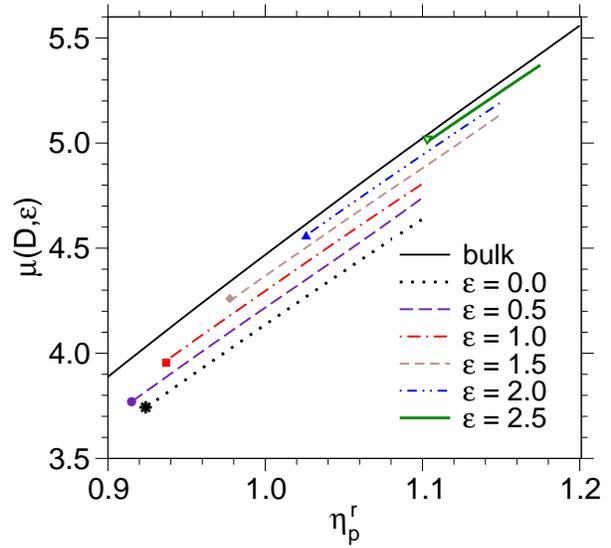}
\caption{Phase diagram of a thin film of colloid-polymer mixtures
with asymmetric walls in the grand-canonical ensemble, choosing
$q=0.8$, $D=5$, with $\eta^{\rm r}_{\rm p}$ as abscissa and $\mu$ as ordinate.
Curves show the coexistence potential $\mu_{\rm coex} (D,\varepsilon)$ of 
the colloids. Full curve denotes the result for
the bulk (note that the bulk critical point, $\eta^{\rm r}_{\rm p, cr}=0.766$ 
is off the scale of this figure). Full symbols mark the
critical points of the films. From De Virgiliis et al.~\cite{63}
\label{fig17}}
\end{center}
\end{figure}

Figure \ref{fig17} shows the phase diagram for $D=5$ and various
choices of $\varepsilon$ in the grand-canonical representation
(this is the counterpart of Fig.~\ref{fig9}a for capillary
condensation, where $D$ was varied for $\varepsilon=0$, while here
we study the variation with $\varepsilon$ at fixed $D$). One can
see that by increasing $\varepsilon$ the coexistence curves of the
thin film move closer towards the bulk coexistence curves, and for
$\varepsilon=2.5$ the deviation from the bulk indeed is very
small, but the critical point is strongly shifted (from
$\eta^{\rm r}_{\rm p,cr} =0.766$ in the bulk to $\eta^{\rm r}_{\rm p,cr}=1.106$
in the thin film).

This behavior is qualitatively similar to what has been found for
the Ising ferromagnet with competing surface magnetic fields
\cite{11,38}, the generic model for which the interface
localization transition was studied for the first time.

Note that the curve for $\varepsilon=0$ in Fig.~\ref{fig17}
represents capillary condensation (and a similar conclusion
applies to the case $\varepsilon=0.5$ as well). Figure \ref{fig17}
implies that varying $\varepsilon$ one can completely smoothly
cross over from capillary condensation-like behavior to interface
localization-like behavior, when $\varepsilon$ is increased. In
view of the qualitative description of Fig.~\ref{fig13} this is
somewhat surprising: in the capillary condensation transition, the
two liquid-vapor interfaces bound to the walls annihilate each
other, the slit pore gets almost uniformly filled with liquid. In
the interface localization transition, one has an interface on
both sides of the transition, it just has jumped at the transition
from one wall to the other.

How can one then reconcile Figs.~\ref{fig13} and~\ref{fig17} with
each other? The clue to the problem is, of course, that the
picture of the states in Fig.~\ref{fig13} is far too simplified,
it ignores the variations of the densities close to the wall.
Therefore the states with ``interfaces bound to walls'' are a
simplification, which lose its meaning when the ``phase'' in
between the interface and the wall can no longer be clearly
identified with bulk-like properties (as is actually the case, see
Fig.~\ref{fig15}). While one can clearly imagine to transform the
left phase diagram of Fig.~\ref{fig13} into the right one by
smooth changes, one should not take the sketches that illustrate
the character of the phases too literally. The failure of these
sketches, however, also means that one must be careful with all
approaches where wetting phenomena and interface localization
transitions \cite{24,25,26,27,36,37,38} are simply described in
terms of the ``interface hamiltonian'' picture, since according to
this description the distance $\ell$ of the interface from the
wall is the single degree of freedom (on a mean-field level) left 
to analyze the problem.

\section{Summary and Outlook}
In this brief review we have emphasized that confinement has very
interesting effects on soft matter systems, both with respect to
the structure and the phase behavior of these systems. Of course,
confinement also has very interesting consequences on the dynamics
of soft matter systems (see e.g.~\cite{13,14,15,16,17,18,19} for
recent discussions), but this aspect has been completely outside
of the focus of our review.

We also have focused on the case which we consider to be the
simplest case, confinement between two flat and ideally parallel
walls a finite distance $D$ apart. Practically more important, of
course, is the confinement in random porous media
\cite{10,12,20,21,22,23}. However, in this case the random
irregularity of the confining geometry is a serious obstacle for a
detailed understanding. There is ample evidence (both from
experiment \cite{126,127} and simulations \cite{128,129,130}) that
the liquid-vapor type phase separation or demixing of binary fluid
mixtures under such confinement is seriously modified, but the
character of this modification has been under discussion since a
long time \cite{126,127,128,129,130,131,132}. De Gennes \cite{131}
argued that due to the random arrangement of the pore walls (which
prefer one of the coexisting phases over the other) the problem
can be mapped to the random field Ising model \cite{133,134}.
While for a long time the existing evidence
\cite{126,127,128,129,130} was inconsistent with this suggestion,
in recent work on colloid-polymer mixtures confined by a fraction
of colloids that are frozen in their positions and do not take
part in the phase separation, evidence for the random field Ising
behavior was obtained \cite{132}. Note that then it is necessary
to reach a regime where the correlation length has grown to a
large enough distance, much larger than the characteristic linear
dimension of the confining particles.

Another case, that has received ample consideration in the
literature (see \cite{12} for further references) but was
disregarded here, is the confinement in a quasi-one-dimensional
cylindrical geometry. In this case again the structure is
typically inhomogeneous in the radial direction perpendicular to
the walls of the cylinder. The correlation length can grow
indefinitely only in one direction, along the cylinder axis,
however. Therefore the phase transition to this laterally
segregated state is a gradual (rounded) transition only, and even
for conditions where in the bulk the system is strongly segregated
(with an interfacial tension $\gamma$ between the coexisting
phases which is not small in comparison with $k_BT$), there is no
macroscopic phase separation possible: rather one predicts that
only phase separation into domains of finite size can occur (cross
sectional area $A$ of the cylinder and length $\ell_{\rm d}$ of the
domains), where \cite{135,136} $\ln \ell_d \propto A \gamma/k_BT$.
This happens because thermal fluctuations prevent the
establishment of true long range order by the spontaneous
generation of transverse interfaces (across the cylinder). This
has the consequence that also, in principle, the capillary
condensation or evaporation transitions in cylindrical geometry
are, in full thermal equilibrium, not perfectly sharp but rounded.
In practice, this effect often is masked by nonequilibrium
phenomena (pronounced hysteresis occurs!) and hence we are not
aware of careful studies where this rounding has been
demonstrated. Note that mean-field treatments \cite{12} miss such
fluctuations effects, of course. Also, when one considers
cylindrical geometries with ```competing walls'' (e.g. a cylinder
with a square cross section, where the upper walls prefer one
phase and the lower walls prefer the other phase \cite{137,138})
the interface localization transition in such a ``double
wedge''-geometry is rounded and not sharp, as simulations show
where one considers the limit that the length of the cylinder gets
macroscopic while its cross section stays finite \cite{138}. Since
recently it has become possible to create artificial cylindrical
nanochannels with diameters between 35 and 150 nm \cite{139}, it
would be interesting to study phase separation in such
nanochannels experimentally as well.

Using the AO model of colloid-polymer mixtures as an example that
is well suited for simulation studies \cite{61,62,63}, we have
discussed simulation evidence for the theoretical concepts on
capillary condensation and interface localization transitions
\cite{11,33,34,35,36,37,38}. In particular, the predictions for
the shift of the critical point have been found to be compatible
with the simulation results, and it was also argued that the
critical behavior of the lateral phase separation in the thin film
has the character of the two-dimensional Ising model (although in
practice one is mostly in a crossover region where ``effective
exponents'' in between the $d=2$ and $d=3$ limits apply, which do
not have a deep theoretical significance). Clearly, it would be
nice to have also experiments that confirm the findings of theory
and simulation on the phase behavior of confined fluid mixtures.

One crucial assumption of the work reviewed here was that the
wetting transitions (that occur in the limit when the film
thickness $D$ diverges to infinity) are of second order \cite{11}.
If first-order wetting occurs, much more complicated phase
diagrams under confinement result \cite{11}. To some extent, this
problem has been worked out for symmetrical polymer blends under
confinement \cite{65,66,67,68,69,70,71,72}, and we refer the
reader to these papers for details. In particular, it also should
be possible to realize situations in between capillary condensation
and interface localization transitions \cite{67}.

Finally, we draw attention to more complicated ordering phenomena
under confinement. A problem that has found a lot of attention is
the effect of confinement in thin films on the blockcopolymer
mesophase ordering \cite{140,141,142,143}. For more or less
symmetric composition of a diblock copolymer, the mesophase
observed in the bulk is a lamellar ordering \cite{57}. The
question that is then discussed in the literature (both
experimentally and theoretically, see \cite{140,141,142} for
further references), is whether the lamellae are oriented parallel
or perpendicular to the confining walls, and transitions in the
number of lamellae that fit into the thin film, etc.~(remember
that the lamella thickness depends on temperature, chain length of
the polymers, and other control parameters \cite{57}). For
asymmetric compositions $A_fB_{1-f}$ of a diblock copolymer,
however, already in the bulk melts other mesophases appear, such
as hexagonal patterns of A-rich cylinders in B-rich background, or
cubic structures, where A-rich cores of micelles form a periodic
lattice in the B-rich background, or vice versa \cite{57}. For
triblock copolymers, many much more complex mesophases occur, and
the question how all this self-assembly of block copolymers is
affected by confinement due to walls is still under both
theoretical and experimental investigation (see e.g.~\cite{143}).

Other very interesting confinement effects in soft matter occur
when orientational order is involved, e.g.~when a colloidal
dispersion undergoing a transition from isotropic to nematic
phases is confined by walls (see e.g.~\cite{144,145,146,147}).
Confinement may enhance the nematic ordering tendency (``capillary
nematization'' \cite{144} is the analog phenomenon of capillary
condensation), but one needs also to take into account the tensor
character of the order parameter of liquid crystals. Thus near a
wall a biaxial character of the ordering occurs even when in the
bulk the ordering is uniaxial. Also the boundary conditions at the
walls can be envisaged such that one wall prefers parallel and the
other wall perpendicular alignment, leading to a tilted structure
of the ordering across the film \cite{145}. These remarks are by
no means intended as an exhaustive discussion, but just want to
draw the attention of the reader to this wealth of interesting problems.

\underline{Acknowledgments}: Support from the Deutsche
Forschungsgemeinschaft (DFG) via SFB TR6/A5 is gratefully
acknowledged. We are indebted to S. Dietrich, R. Evans, H.
Lekkerkerker, H. L\"owen, M. M\"uller, M. Schmidt and P. Virnau
for many stimulating discussions.

\clearpage

\end{document}

\maketitle
\renewcommand{\thefootnote}{\fnsymbol{footnote}}

\noindent Insert abstract here. It summarises the content of the
article and should be kept a single paragraph.\section{Introduction}
\label{intro}{The body text should appear here. Headings and subheadings should follow the allowed hierarchy: for example, B headings should only be used in a section with an A heading, and D headings should follow from a C Heading. A headings usually consist of the major sections in an article, {\it e.g.} Introduction, Experimental, Results and Discussion.\\

Please remember to input section labels for easy
cross-referencing.}

\section{Experimental}
\label{exper}
\begin{figure}[ht]
\begin{center}
\includegraphics{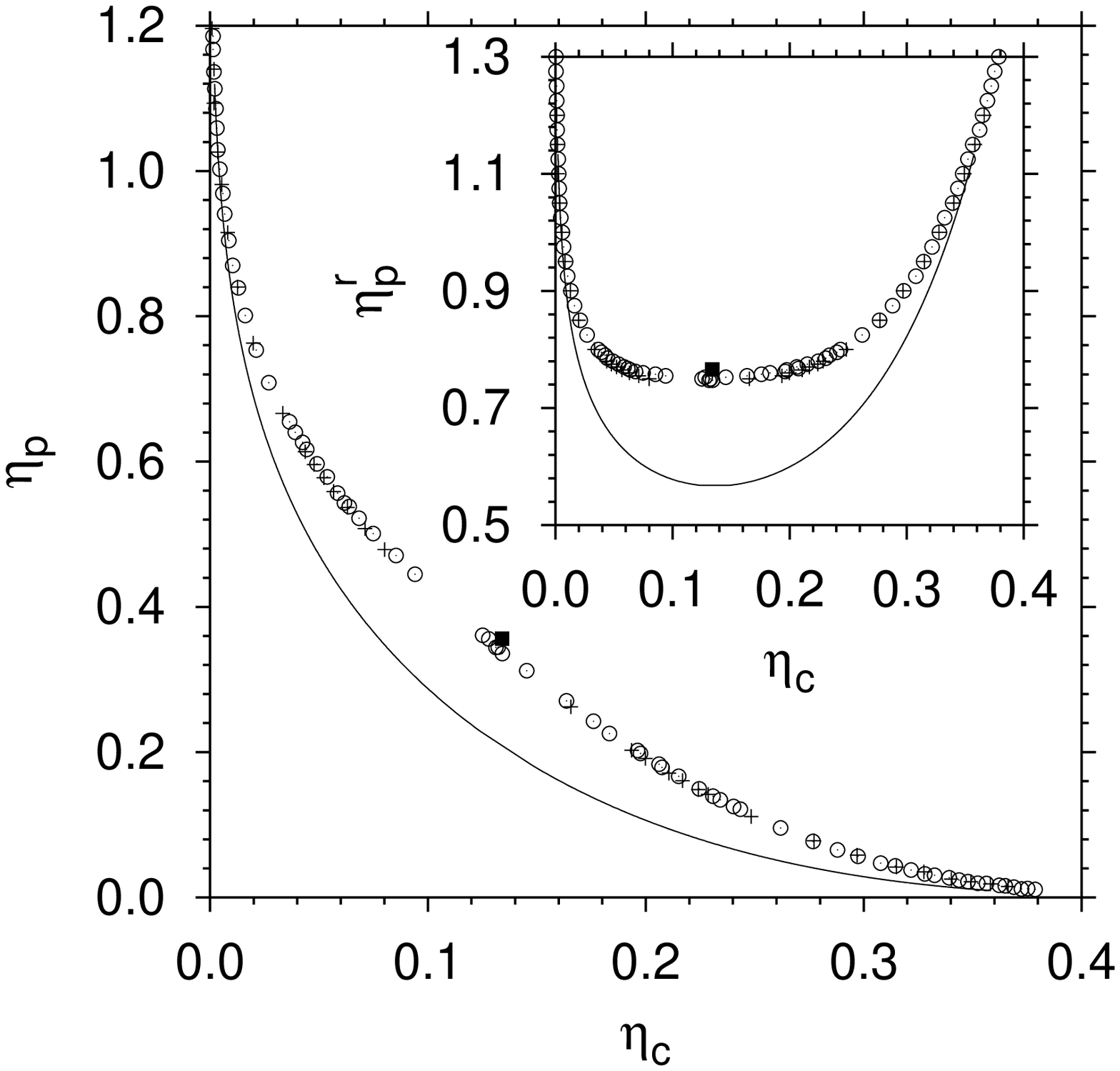}
\caption{Caption of Fig. 1.}
\label{fig1}
\end{center}
\end{figure}

Include in the text also the figures, e.g.,
Fig~\ref{fig1}

\subsection{Reagents and apparatus}
\label{reag_app}Please feel free to use any style-files for enhancing the output, as long as those packages are available on CTAN. Otherwise, they will have to be submitted together with the main file.\\

When using footnotes, please follow the command \verb|\footnote[X]{The footnote text}| where for X the numbers from 1 to 9 should be input in succession.\\

\noindent{\bf 2.1.1 C Heading.} Please use this format for C headings...\\

\noindent{\it 2.1.1.1 D Heading.} And this one for D headings

%The references should start on their own page.
\vspace{1.5cm}
\noindent \textbf{Author Name 1,$^a$ Author 2 Name$^b$ and Author 3 Name$^a$$^b$}\\
\noindent $^a$ \textsl{Address, Address, Town, Country. E-mail: xxxx@aaa.bbb.ccc\\
\noindent $^b$ Address, Address, Town, Country. E-mail:
xxxx@aaa.bbb.ccc}

\clearpage \onecolumn
The tables should be submitted normally
after the reference list, starting on a separate page.

\begin{table}

\begin{center}

\caption{\label{tab1}Insert table caption here}

\begin{tabular}{llll}

\hline

Header 1 & Header 2 & \multicolumn{2}{l}{Header 3}\\ \cline{3-4}

&&Subheader 1 & Subheader 2 \\ \hlineColumn 1 & Column 2 & Column 3 & Column 4\\Column 1 & Column 2 & Column 3 & Column 4\\

\hline

\end{tabular}

\end{center}

\end{table}

%Please compile a list of all figure captions on a separate page:
\clearpage
\begin{list}{}{\leftmargin 2cm \labelwidth 1.5cm \labelsep 0.5cm}

\item[\bf Fig. 1] Caption of Figure 1.
\item[\bf Fig. 2] Caption
of Figure 2.

\end{list}

\clearpage

